\documentclass[twocolumn,aps,showpacs,showkeys,floatfix]{revtex4}
\usepackage{amssymb,amsmath,graphics,epsfig}
\usepackage[T1]{fontenc}
\usepackage[latin1]{inputenc}

\renewcommand{\epsilon }{\varepsilon}
\renewcommand{\phi}{\varphi}
\renewcommand{\tilde}{\widetilde}
\renewcommand{\hat}{\widehat}

\newcommand{\LL}{\mathcal{L}}

\newcommand{\NN}{\mathcal{N}}
\newcommand{\OO}{\mathcal{O}}

\newcommand{\SA}{\circ}
\newcommand{\SB}{\bigtriangleup}
\newcommand{\SC}{+}
\newcommand{\SD}{\times}
\newcommand{\SE}{\diamond}
\newcommand{\SF}{\bigtriangledown}

\newcommand{\real}{\mathbb{R}}

\newcommand{\HALF}{\frac{1}{2}}
\newcommand{\FOUR}{\frac{1}{4}}

\def\sigf{\sigma_f}
\def\ga{\gamma}

\begin{document}

\title{
Collective synchronization in populations of globally coupled phase
oscillators with drifting frequencies}

\author{Jacques Rougemont}
\affiliation{Vital-IT, Swiss Institute of Bioinformatics, CIG-UNIL, 1015 Lausanne, Switzerland}
\email{jacques.rougemont@isb-sib.ch}
\author{Felix Naef}
\affiliation{Swiss Institute for Experimental Cancer Research (ISREC), Ecole Polytechnique F\'ed\'erale de Lausanne (EPFL) and Swiss Institute of Bioinformatics, 1015 Lausanne, Switzerland}
\email{felix.naef@isrec.ch}
                                                                               
\begin{abstract}
We generalize the Kuramoto model for coupled phase
oscillators by allowing the frequencies to drift in time
according to Ornstein-Uhlenbeck dynamics.
Such drifting frequencies were recently measured in cellular
populations of circadian oscillator and inspired our work.
Linear stability analysis of
the Fokker-Planck equation for an infinite population is amenable to
exact solution and we show that the incoherent state is unstable
passed a critical coupling strength $K_c(\ga, \sigf)$, where $\ga$ is
the inverse characteristic drifting time and $\sigf$ the asymptotic
frequency dispersion. Expectedly $K_c$ agrees with the noisy Kuramoto
model in the large $\ga$ (Schmolukowski) limit but increases
slower as $\ga$ decreases.
Asymptotic expansion of the solution for $\ga\rightarrow 0$
shows that the noiseless Kuramoto model with Gaussian frequency
distribution is recovered in that limit.
Thus varying a single parameter allows to interpolate smoothly between two regimes:
one dominated by the frequency dispersion and the other by phase diffusion.
\end{abstract}

\keywords{synchronization, phase oscillators, phase locking, Kuramoto model}
\pacs{05.40.-a, 05.45.Xt, 89.75.-k, 87.19.Jj}
\maketitle

\section{Introduction}

Synchronization phenomena occur recurrently in 
physical, chemical and biological systems.
Few noteworthy examples include superconducting currents
in Josephson junction arrays
\cite{strogatz_prl_1996,strogatz_pre_1998},
emerging coherence in populations of chemical oscillators
\cite{kiss},
or the accuracy of central circadian pacemakers in insects and vertebrates
\cite{reppert_97,winfree,sync_book}.
The latter serve as biomolecular time-keeping
devices, which most organisms have evolved to coordinate their
physiology and metabolic
activities with the geophysical light-dark and temperature cycles
\cite{schibler_naef}.

The present work was motivated by recent experiments in mammalian
cell cultures in which the levels of
proteins implicated in the circadian (with $\sim 24$hr periods)
clockwork were monitored using fluorescent reporters
\cite{nagoshi, welsh, Carr}.
It was demonstrated that
individual cellular oscillators generate self-sustained rhythms
in protein abundance
and that populations can be synchronized by treatment with a short
serum shock or light pulse.
Importantly frequencies of individual oscillators are not strictly
constant but drift in time
(see for example Fig.~S2 in \cite{Carr} showing the results in the zebrafish).
Several chemical kinetics models thought to
capture the biochemistry responsible for generating
oscillations in living cells
were shown to exhibit oscillatory instabilities and limit-cycles
\cite{goldbeter, leibler}.
Experimental and theoretical evidence therefore supports a
description of oscillator populations in terms of phase variables.

Our understanding of the onset of collective synchronization in coupled
nonlinear oscillator models has greatly benefited from a large body of
work on the Kuramoto model
\cite{kuramoto_84,Sakaguchi_kuramoto_86,kuramoto_nishikawa_87,strogatz_prl_1992,stogatz_physicaD_2000}
\begin{eqnarray}
\label{kuramoto}
\dot\phi_i=f_i+\frac K N \sum_{j=1}^N\sin(\phi_j-\phi_i)+\xi_i(t)~,
\end{eqnarray}
describing the phase dynamics in a set of weakly coupled identical
non-linear oscillators.
Here, $\phi_i(t)$ represent the phase of the
$i$-th oscillator at time $t$,
and $\xi_i(t)$ are white noise sources with
expectation and covariance
\begin{eqnarray*}
{\rm E}[\xi_i(t)]&=&0~,\\
{\rm Cov}[\xi_i(s),\xi_j(t)]&=&2D\delta _{ij}\delta(s-t)~.
\end{eqnarray*}
%$\langle\xi_i(s)\xi_j(t)\rangle=2D\delta _{ij}\delta(t-s)$ 
The frequencies $f_i$ are static and taken from a distribution
$g(f)$ symmetric around $\mu_f$. Eq.~\ref{kuramoto} is an effective
model for the phase degrees of freedom in a population of
limit-cycle oscillators and assumes a regime where phase and amplitude
dynamics decouple.
The parameter $K$ measures the strength of the all-to-all coupling.
Most exact results are given for the coupling function
$U(\phi_j-\phi_i) = \sin(\phi_j-\phi_i)$. More general interactions
lead to much greater
analytical complexity and were investigated in
\cite{crawford_prl95,crawford_physicaD_1999}.
Critical properties of the model are conveniently studied using the complex
order parameter $R(t)e^{i\psi(t)}=\frac{1}{N}\sum_{j=1}^Ne^{i\phi_j(t)}$
so that collective synchronization occurs
when $R_\infty=\lim_{T\rightarrow\infty}\frac 1 T\int_0^T\! R(t)\,dt$ remains positive
in the infinite population limit.
For the sine coupling model a bifurcation occurs at
$K_c = 2/\int\frac D {D^2+f^2}g(f+\mu_f)\,\!df$
at which the incoherent desynchronized state $R_\infty=0$ 
becomes unstable and a macroscopic number of oscillators phase lock
to the average phase $\psi(t)=\mu_f\,t$ \cite{crawford_jsp_1994,Sakaguchi,strogatz_jsp_1991}.
For $D\rightarrow 0$ the classical Kuramoto result
$K_c=\frac 2{\pi g(\mu_f)}$ \cite{kuramoto_book} is recovered.
Below the critical coupling, the incoherent state is linearly stable when $D>0$ \cite{strogatz_jsp_1991}
but only neutrally stable when $D=0$ with $R(t)$ still decaying to zero \cite{strogatz_prl_1992}.

\section{The Model}

To study the effects of the reported drifts on collective synchronization,
we generalize the Kuramoto model by introducing
a second time scale $1/\ga$ (besides $1/\sigma_f$)
characterizing the frequency drifts.
The frequency dynamics is formulated as an Ornstein-Uhlenbeck (O-U) process
while the phases are coupled following the canonical
all-to-all sine interaction. The model for $N$ oscillators reads
\begin{eqnarray}
\label{model}
\dot f_i(t)&=& -\ga(f_i(t)-\mu_f)+\eta_i(t)~,\\
\dot \phi_i(t)&=& f_i(t)+\frac K N \sum_{j=1}^N\sin\bigl(\phi_j(t)-\phi
_i(t)\bigr)~,\nonumber
\end{eqnarray}
where $\mu_f$ is the average frequency chosen identical
for each oscillator
%(by a change of variable, $\mu_f$ can be assumed to be $0$, see below).
We assume that the $\eta_i$ are independent and identically
distributed white noise sources with
%$\langle\eta_i(s)\eta_j(t)\rangle=2\sigf^2\ga\delta _{ij}\delta(t-s)$.
%%%$\langle\eta_i(s)\eta_j(t)\rangle=\eta^2\delta _{ij}\delta(t-s)$.
\begin{eqnarray*}
{\rm E}[\eta_i(t)]&=&0~,\\
{\rm Cov}[\eta_i(s),\eta_j(t)]&=&\eta^2\delta _{ij}\delta(s-t)~.
\end{eqnarray*}
The solution for $f_i$ 
%$f_i(t)=\mu_f(1-e^{-\ga t})+e^{-\ga t}f_i(0)
%+\int_0^te^{-\ga(t-s)}d\eta_i(s)$
is a Gaussian process with mean and covariance 
\begin{eqnarray}
%\langle f_i(t)\rangle&=&\mu_f+e^{-\ga t}(f_i(0)-\mu_f)~,\\
%\langle f_i(s)f_j(t)\rangle&=&\sigf^2\delta_{ij}\left(e^{-\ga|t-s|}+e^{-\ga(t+s)}\right)~.
{\rm E}[f_i(t)]&=&\mu_f+e^{-\ga t}(f_i(0)-\mu_f)~,\nonumber\\
{\rm Cov}[f_i(s),f_j(t)]&=&\frac{\eta^2}{2\gamma}\delta_{ij}
\left(e^{-\ga|t-s|}-e^{-\ga(t+s)}\right)\nonumber\\
%&=&\frac{\eta^2}{\gamma^2}\delta_{ij}
%\left(\frac \ga2 e^{-\ga|t-s|}+o(\frac1{\ga t})\right)\\
&\stackrel{\gamma t\gg1}{\longrightarrow}&\frac{\eta^2}{\gamma^2}\delta_{ij}\delta(t-s)~.
%{\rm Cov}[f_i(t),f_j(t)]&=&\frac{\eta^2}{2\gamma}\delta_{ij}
\end{eqnarray}

In the following we use as independent parameters the
asymptotic frequency dispersion $\sigma_f^2=\frac{\eta^2}{2\gamma}$
and the damping $\gamma$, which are in principle both accessible
experimentally.
Then,
$
{\rm Cov}[f_i(s),f_i(t)]\rightarrow2\sigma_f^2/\gamma\,\delta(t-s)
$ when $\gamma t\gg1$~. 
To remind the significance of this regime we note
for $K=0$ the phases $\phi_i(t)$ also follow Gaussian processes with
$
%\begin{eqnarray*}
{\rm Cov}[\phi_i(t),\phi_i(t)]
%\frac{2\sigma_f^2}{\ga}\int_0
%^t\int_0^t\left(\delta(s-s')+o(\frac1{\ga s})\right)ds\,ds'\\
\rightarrow\frac{2\sigma_f^2}{\gamma}t
%\end{eqnarray*}
$
asymptotically for $\gamma t\gg 1$.
Because of the linear time dependence, this regime (Schmolukowski) describes
phase diffusion with constant $D=\frac{\sigma_f^2}\gamma$.

Although it is {\it a priori} unclear whether this model exhibits
a bifurcation, we expect that the large $\ga$ behavior
reminiscent of phase diffusion will converge
to the Kuramoto model (Eq.~\ref{kuramoto}) with a
frequency distribution given by
$g(f)=\delta(f-\mu_f)$ and white noise
strength $D=\sigf^2/\ga$, and thus exhibit a bifurcation at $K_c=2D$.
However the small $\ga$ behavior is less obvious
since we are simultaneously concerned with long time properties.
For fixed $\sigma_f$, we anticipate that synchronization should be hardest
for strictly static oscillators ($\gamma=0$).
This case corresponds to the noiseless $D=0$ Kuramoto
model with Gaussian frequency dispersion
$g(f)=\NN_{\mu_f,\sigf}(f)\equiv(\sqrt{2\pi}\sigf)^{-1}e^{-(f-\mu_f)^2/2\sigf^2}$, 
so that $K_c=2\sqrt{2/\pi}\,\sigf$.
As the frequency dynamics loses stiffness (when $\gamma$ increases),
we expect the synchronization threshold to be facilitated
by the frequency drifts.

We study the infinite population model $N\rightarrow\infty$ by formulating
a Fokker-Planck equation for the time
dependent joint density $p(\phi, f, t)$.
The all-to-all interaction term
\begin{equation}
\label{roft}
\frac K N \sum_{j=1}^N\sin\bigl(\phi_j(t)-\phi _i(t)\bigr)
\,=\,KR(t)\sin(\psi(t)-\phi_i(t))
\end{equation}
has well known mean-field character and can be replaced
for $N\rightarrow\infty$
by
\begin{equation}
\label{ergodic}
K\int_0^{2\pi}\!d\theta\int\!dg\,p(\theta,g,t)\sin(\theta-\phi_i(t))~.
\end{equation}
We obtain
\begin{eqnarray}
\label{fokker}
\frac{\partial p}{\partial t}&=&
\ga\sigma_f^2\frac{\partial^2p}{\partial f^2}
-f\frac{\partial p}{\partial \phi}
+\ga(f-\mu_f)\frac{\partial p}{\partial f}+\ga p\\
&&-K\frac{\partial \bigl(c(p, \phi)\,p\bigr)}{\partial \phi}\nonumber~.
\end{eqnarray}
This is the known expression for an O-U process augmented by a phase
coupling involving $c(p,
\phi)=\int_0^{2\pi}\!\!d\theta\int\!dg\,p(\theta,g,t)\sin(\theta-\phi)$,
%\begin{eqnarray}
%$$
%c(p, \phi)%\,=\,R(p)\sin(\psi(p)-\phi)%\nonumber\\
%\,=\,\int_0^{2\pi}\!\!\!\!d\theta\int\!dg\,p(\theta,g,t)\sin(\theta-\phi)~,
%$$%\end{eqnarray}
which makes Eq.~\ref{fokker} non-linear as a consequence of Eq.~\ref{ergodic}.

\section{Stability Analysis}

We next discuss the linear stability of the incoherent
stationary solution $p_0(\phi, f)=\NN_{\mu_f,\sigf} (f)/(2\pi)$ in first
order. For reasons that will become clear, we factorize a term 
$\NN^{1/2}_{\mu_f,\sigf}(f)$
off the perturbation and write
$$
p(\phi, f, t)=p_0(f)+\NN^{1/2}_{\mu_f,\sigf}(f)
\epsilon\bigl(\phi-\mu_ft,\sigf^{-1}(f-\mu_f),\gamma t\bigr)~,
$$
where $\epsilon(\tilde\phi, \tilde f, \tilde t)$ is a small perturbation
expressed in a rotating frame using rescaled frequency and time variables.
By plugging this ansatz into Eq.~\ref{fokker} we obtain the linearized problem
$\frac{\partial\epsilon}{\partial t}=\LL\epsilon+\OO(\epsilon^2)$
where $\LL$ is the linear operator
\begin{eqnarray*}
%\label{linearpart}
\LL\epsilon&=&\frac{\partial^2\epsilon}{\partial f^2}
-\frac{\sigf}{\gamma}f\frac{\partial\epsilon}{\partial \phi}
+\left(\HALF-\FOUR f^2\right)\epsilon\hskip 2.5cm\\
\lefteqn{+\frac{K}{2\pi\gamma}\NN^{1/2}_{0,1}(f)
\int_0^{2\pi}\!\!\!\!d\theta\!
\int\!dg\,\NN^{1/2}_{0,1}(g)\epsilon(\theta,g,t)\cos(\theta-\phi)~.}
%\nonumber
\end{eqnarray*}
Decomposing $\epsilon$ as a Fourier series in $\phi$, 
$\epsilon (\phi,f,t)\,=\,\sum_{n=-\infty}^\infty\epsilon_n(f,t)e^{-in\phi}$,
we obtain for the coefficients $\epsilon_n$
\begin{eqnarray}
\label{fourier}
\frac{\partial\epsilon _n}{\partial t}
&=&\frac{\partial^2\epsilon _n}{\partial f^2}
+\left(\HALF-\FOUR f^2+\frac{in\sigf}{\gamma}f\right)\epsilon_n\nonumber\\
&&+\delta_{1|n|}\frac{K}{2\gamma}
\NN^{1/2}_{0,1}(f)\int\!dg\,\NN^{1/2}_{0,1}(g)\epsilon_n(g,t)\nonumber\\
&\equiv&\LL_n\epsilon_n
+\delta_{1|n|}\frac{K}{2\gamma}\langle\epsilon _n,\NN^{1/2}_{0,1}\rangle\NN^{1/2}_{0,1}~.
%&\equiv&\left(\LL_0+\frac{in\sigf}{\gamma}f\right)\epsilon_n
%+\delta_{1|n|}\frac{K}{2\gamma}\langle\epsilon _n,\NN^{1/2}_{0,1}\rangle\NN^{1/2}_{0,1}~.\quad~
\end{eqnarray}
We notice that the first term representing the frequency dynamics resembles
the harmonic oscillator plus a complex part, which
can be removed by applying
the translation operator $U_\theta$ defined by
$\bigl(U_\theta f\bigr)(x)=f(x-\theta)$. We note that 
$\LL_n=U_{2in\sqrt{a}}\hat\LL_nU_{-2in\sqrt{a}}$, 
where we have set $a=(\sigf/\gamma)^2$ and the operator 
$\hat\LL_n=\partial_f^2+\HALF-n^2a-\FOUR f^2$ is self-adjoint on
$L^2(\real)$ and
has pure point spectrum
$\Sigma(\hat\LL_n)=\{\lambda_{n\ell}=-\ell-n^2a\,:\,\ell=0,1,2,\dots\}$.
Its eigenfunctions are given in terms of the Hermite
functions \cite{reed-simon}
\begin{eqnarray*}
H_0(x)&=&\pi^{-\FOUR}e^{-\HALF x^2}~\mbox{and}~\\
H_\ell(x)&=&(2^\ell\ell!\sqrt{\pi})^{-\HALF}(-1)^\ell
e^{\HALF x^2}\partial_x^\ell e^{-x^2}~,~\ell\,=\,1,2,\dots
\end{eqnarray*}
as follows:
%\begin{equation}
%\label{eigval}
$$
\hat\LL_n\Phi_\ell\,=\,\lambda_{n\ell}\Phi_\ell~,\quad
\Phi_\ell(f)\,=\,2^{-1/4}H_\ell(2^{-1/2}f)~.\\
$$%\end{equation}
Therefore $\{U_{2in\sqrt{a}}\Phi_\ell:\ell=0,1,2,\dots\}$ forms an
orthonormal family which diagonalizes $\LL_n$. 
Notice that the largest eigenvalue for each $n$ is
$\lambda_{n0}=-n^2a$.

Linear stability follows directly except for $|n|\neq 1$ and
$K>0$. Indeed, 
for $n=0$, we find $\lambda_{00}=0$ with corresponding
eigenfunction is $\Phi_0=\NN^{1/2}_{0,1}$.
However, this function lies outside the space of relevant perturbations
because the normalization of $p$, 
$\int_0^{2\pi}d\phi\,\int df\,p(\phi,f,t)=1$,
requires orthogonality of $\NN^{1/2}_{0,1}$ and $\epsilon_0(f,t)$
through $\int\NN^{1/2}_{0,1}\epsilon_0(f,t)df=0$.
Subsequent eigenvectors have negative eigenvalues.
For all other $|n|\neq 1$
the coupling term in Eq.~\ref{fourier} vanishes
and the incoherent state $p(\phi, f, t)=p_0(f)$ 
is linearly stable as a consequence of the
strictly negative spectrum of $\LL_n$.
The same holds for all $n$ in the absence of coupling $K=0$.
%%%%%%%%%%%%%%%%%%%%%%%%%%%%%%%%%%%%%%%
\setlength{\unitlength}{1mm}
\begin{figure}
%\centering
\includegraphics[scale=0.8]{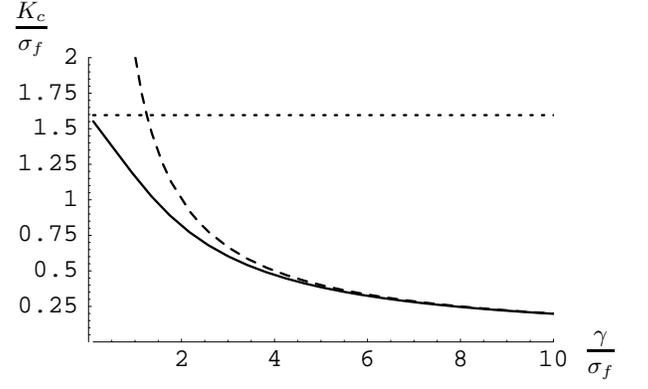}
\put(1,2){$\displaystyle\frac{\gamma}{\sigma_f}$}
\put(-75,45){$\displaystyle\frac{K_c}{\sigf}$}
\caption{Behavior of $K_c$ as a function of $\ga$ as given by Eq.~\ref{kcritical} with
$a=(\sigf/\gamma)^2$ (continuous line). 
Dashed line represents the Kuramoto model with
identical frequencies and $K_c=2D=2\sigma_f^2/\ga$. The
$\gamma\rightarrow 0$ limit reproduces the $\ga=0$ model with Gaussian
frequency dispersion and gives $K_c/\sigma_f=2\sqrt{2/\pi}=1.5957$ (dotted 
line).}
\label{fig1}
\end{figure}
%%%%%%%%%%%%%%%%%%%%%%%%%%%%%%%%%%%%%%%

For the remaining case $n=\pm 1$ and $K>0$,
we notice that the coupling term
in Eq.~\ref{fourier} also vanishes for all directions
orthogonal to $\NN^{1/2}_{0,1}$,
leaving a one-dimensional space that could develop an instability.
We write the eigenvalue problem for Eq.~\ref{fourier} implicitly as
\begin{eqnarray*}
\LL_n\epsilon_n
+\delta_{1|n|}\frac{K}{2\gamma}\langle\epsilon_n,\NN^{1/2}_{0,1}\rangle
\NN^{1/2}_{0,1}=\lambda\,\epsilon_n~.
\end{eqnarray*}
Using the resolvent equation 
\begin{eqnarray*}
(\lambda-\LL_n)^{-1}&=&(\lambda-U_{2in\sqrt{a}}\hat\LL_nU_{-2in\sqrt{a}})^{-1}\\
&=&U_{2in\sqrt{a}}(\lambda-\hat\LL_n)^{-1}U_{-2in\sqrt{a}}
\end{eqnarray*}
we obtain
\begin{eqnarray*}
\epsilon_n&=&\frac{K}{2\gamma}\langle\epsilon _n,\NN^{1/2}_{0,1}\rangle
U_{2in\sqrt{a}}(\lambda-\hat\LL_n)^{-1}U_{-2in\sqrt{a}}\NN^{1/2}_{0,1}\\
&=&\frac{K}{2\gamma}\langle\epsilon _n,\NN^{1/2}_{0,1}\rangle
\sum_{j=0}^\infty
\frac{\langle\Phi_j,U_{-2in\sqrt{a}}\NN^{1/2}_{0,1}\rangle}{\lambda-\lambda_{nj}}
U_{2in\sqrt{a}}\Phi_j~,
\end{eqnarray*}
where we have used the spectral decomposition
$\hat\LL_nf=\sum_j\lambda_j\langle\Phi_j,f\rangle\Phi_j$.

To find the critical coupling $K_c$ above which the incoherent state becomes
linearly unstable, we need to monitor when the
largest eigenvalue crosses the imaginary axis.
After projecting onto $\NN^{1/2}_{0,1}$, simplifying the factors 
$\langle\epsilon _n,\NN^{1/2}_{0,1}\rangle$ on both sides of the
equation, and setting $\lambda=0$ we
find an equation for $K_c$:
\begin{eqnarray}
\frac{2\gamma}{K_c}&=&
\sum_{j=0}^\infty
\frac{\langle\Phi_j,U_{-2i\sqrt{a}}\NN^{1/2}_{0,1}\rangle^2}{-\lambda_{1j}}
\,=\,e^a\sum_{j=0}^\infty\frac{(-a)^j}{j!(j+a)}\nonumber\\
&=&e^aa^{-a}\!\int_0^a\!\!t^{a-1}e^{-t}dt
\,=\,e^aa^{-a}\gamma(a,a)~,
\label{kcritical}
\end{eqnarray}
where $\gamma(a,x)$ is the lower incomplete $\Gamma$-function.

The behavior of $K_c$ together with the Kuramoto model asymptotes for
$\ga\rightarrow\infty$ and $\ga\to0^+$ limit are shown in Fig.~\ref{fig1}.
It is noticeable that we find a bifurcation for all values of $\gamma$.
$K_c$ strictly decreases from a finite
$\gamma=0$ as $\gamma$ increases, asymptotically behaving as
$K_c=2\sigma_f^2/\ga$. 
The analytical result thus supports the following picture:
for small $\gamma$, the dominant source of fluctuations against which
the coupling must work to achieve synchronization is the
(Gaussian) frequency dispersion. As $\gamma$ increases while $\sigma_f$
is kept fixed, faster frequency drifts help synchrony by preventing
individual oscillators with detuned frequency to stay out of tune for
too long. Indeed with drifting frequencies every individual oscillators
fluctuates around the mean frequency $\mu_f$ with a time scale $\gamma^{-1}$.
In the large $\gamma$ regime, the effective frequency dispersion vanishes
and the coupling force needs to synchronize noisy but otherwise identical frequency
oscillators. As predicted by the phase diffusion limit, the effective white
noise strength $D$ and hence $K_c$ decrease as $\gamma^{-1}$.

We now discuss the asymptotic regimes in detail: 
the small $\gamma$ limit follows from
reverting to the original variables and using the asymptotic expansion of
$\gamma(a,a)$ 
(using Stirling's formula and \cite{abramowitz}:
6.5.3, 6.5.22, and 6.5.35). We obtain in the
limit $\gamma\to0^+$
$$
2\left(\frac {K_c} \sigf\right)^{-1}\,=\,
\sqrt{\frac{\pi}{2}}+
\frac 1 3 \frac{\gamma}{\sigf}+\frac{\sqrt{2\pi}}{24}\frac{\gamma^2}{\sigf^2}+\OO(\gamma^3)~.
$$
This proves that the model continuously interpolates to the noiseless ($D=0$)
model and that the $\gamma\rightarrow 0$
recovers the $\gamma=0$ transition predicted in the original
Kuramoto model at $K_c/\sigma_f=2\sqrt{2/\pi}$.
In the opposite regime $\ga\to\infty$ (thus $a\to 0$) we find
(\cite{abramowitz} 6.5.12, 13.1.2)
$$
a^{-a}e^a\gamma(a,a)\,=\,a^{-1}M(1,1+a,a)\,\sim\,a^{-1}\bigl(1+\OO(a)\bigr)~,
$$
where $M(\cdot,\cdot,\cdot)$ is the confluent hypergeometric
function. This leads $K_c\sim2\sigf^2/\ga+\OO(\ga^{-2})$
and hence proves the convergence to the white noise model
(Eq.~\ref{kuramoto}) with $D=\sigf^2/\ga$.

Finally, we mention a generalization that
includes a white noise source in the phase equation (as in Eq.~\ref{kuramoto})
in addition to the correlated frequency fluctuations.
This leads an additional diffusion term
$
-D\frac{\partial^2p}{\partial \phi^2}
$
in Eq.~\ref{fokker}.
Following the steps above readily extends Eq.~\ref{kcritical} to
$$
\frac{2\gamma}{K_c}\,=\,\,e^aa^{-(a+b)}\gamma(a+b,a)~,
$$
where $b=D/\gamma$, with similar qualitative behavior.
In particular, $K_c$ asymptotes to $2(\sigma_f^2/\gamma+D)$ for large
$\gamma$ and has finite $\gamma\rightarrow 0$ limit.

\section{Numerical Simulations}

We have performed numerical simulations of Eq.~\ref{model} to explore
the behavior of $R(t)$ (see Eq.~\ref{roft}) and in particular
$R_\infty$ in function of the reduced
coupling $K_r=(K-K_c)/K_c$.
To verify the analytical results and study the scaling $R_\infty=\kappa K_r^\beta$
above the bifurcation, we simulated a finite number of oscillators
using the exact solution for the frequency part, leading to the updates
$f_i(t+\!dt)=f_i(t)\,e^{-\ga\,\!dt} + \mu_f(1-e^{-\ga\,\!dt})+
\eta\,\sigf\sqrt{1-e^{-2\ga\,\!dt}}$
and $\phi_i(t+\!dt)=\phi_i(t) + (f_i(t)+\frac K N \sum_j\sin(\phi_j-\phi_i))\,\!dt$
where $\eta$ is a Gaussian random number.
We used Eq.~\ref{roft} to compute $R(t)$ and transients were removed by waiting
until the solutions from two different initial conditions
$\phi_i(t=0)=0$ and $\phi_i(t=0)$ taken randomly converged to the same trajectory.
The steady state value $R_\infty$ was subsequently estimated by averaging $R(t)$ over time.

\begin{figure}
\centering
\includegraphics[scale=0.4,angle=270]{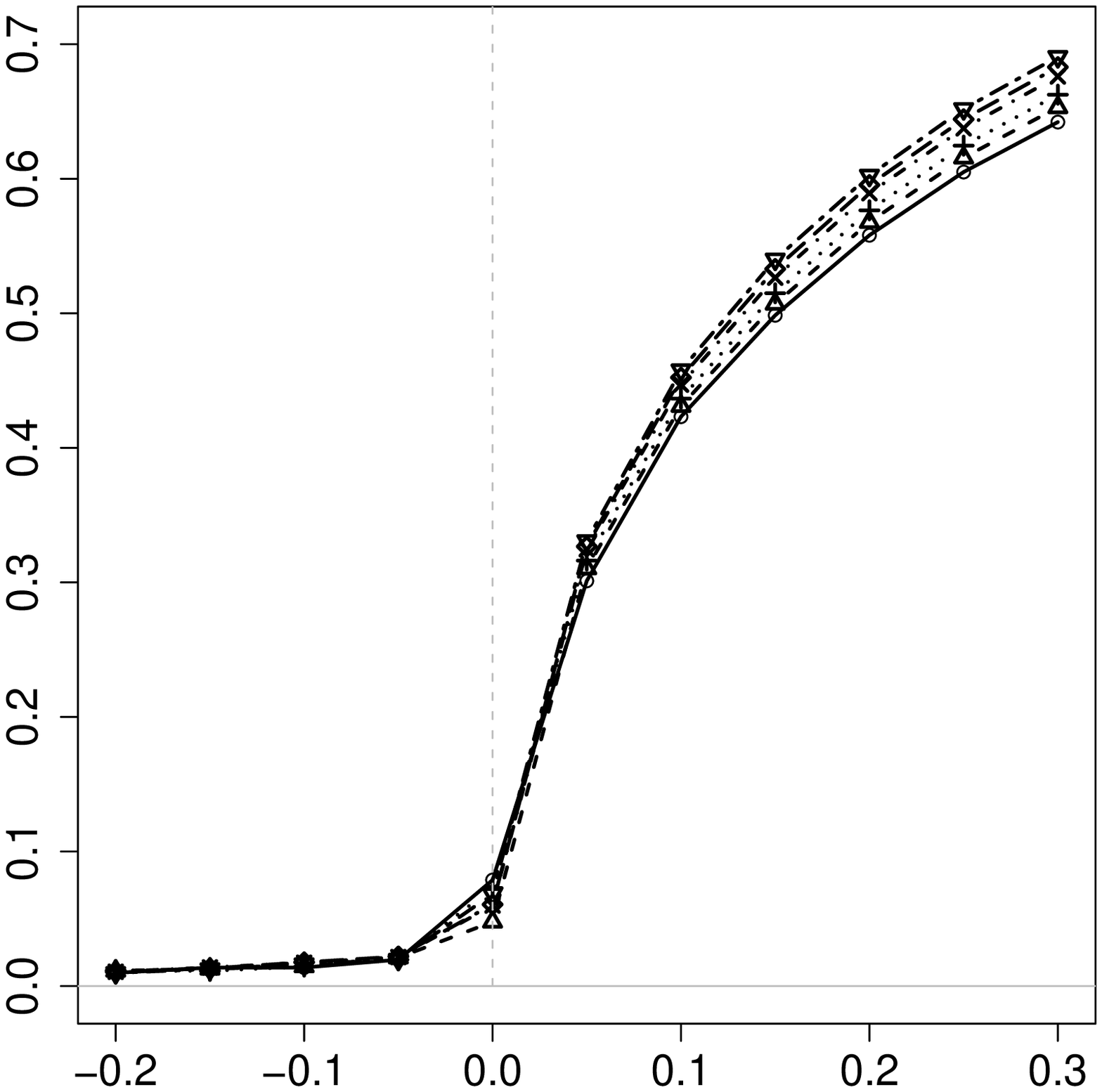}
\put(-58,-1){\includegraphics[scale=0.17,angle=270]{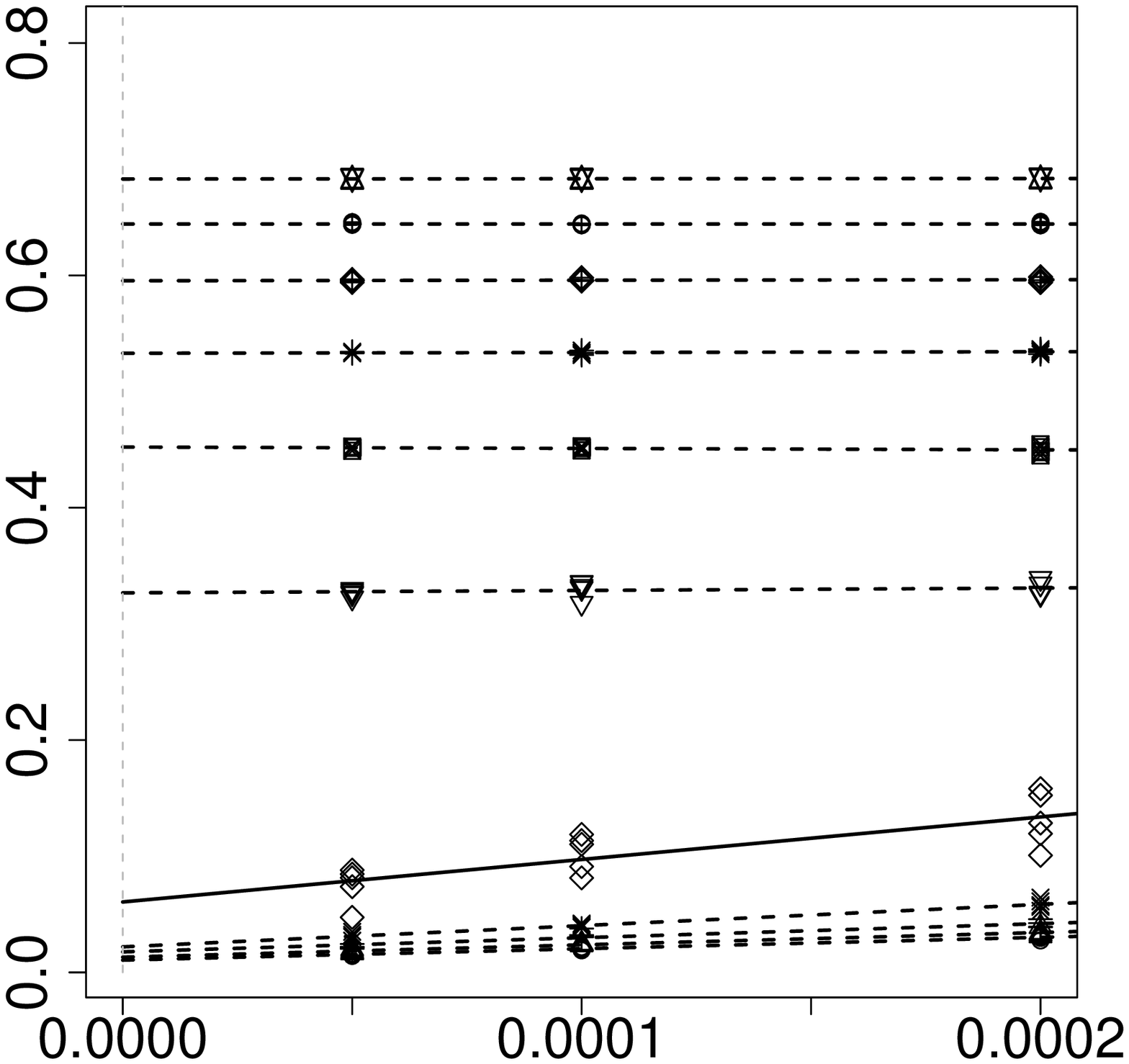}}
\put(-70,-31){$R_\infty$}
\put(-39,-68){$(K-K_c)/K_c$}
\put(-60,-12){\tiny$R_\infty$}
\put(-45,-31){\tiny$1/N$}
\caption{Numerical simulation of Eq.~\ref{model}. 
Estimation of $R_{\infty}$ was obtained using finite size
scaling for systems of sizes $N=5000$, $10000$ and
$20000$. Eq.~\ref{kcritical} was used for $K_c$ to set the reduced coupling
$K_{r}=(K-K_c)/K_c$. Values for $\gamma$ were $4(\SA)$,
$3(\SB)$, $2.5(\SC)$,
$2(\SD)$, $1.8(\SE)$ and $1.6(\SF)$
and $\sigf=1$. In each simulation, $10^5$ time steps of size
$dt=0.01$ were performed. We verified that the dependence in the step size was
weak. Inset: $1/N$ finite size
scaling for $\ga=1.8$. $K_r=0$ is the solid line, smaller (resp. larger)
$K_r$ are below (resp. above) $K_r=0$.
The extrapolated value for $1/N=0$ is used in the main panel.}
\label{kcfig}
\end{figure}

\begin{figure}
\centering
\includegraphics[scale=0.32,angle=270,origin=br]{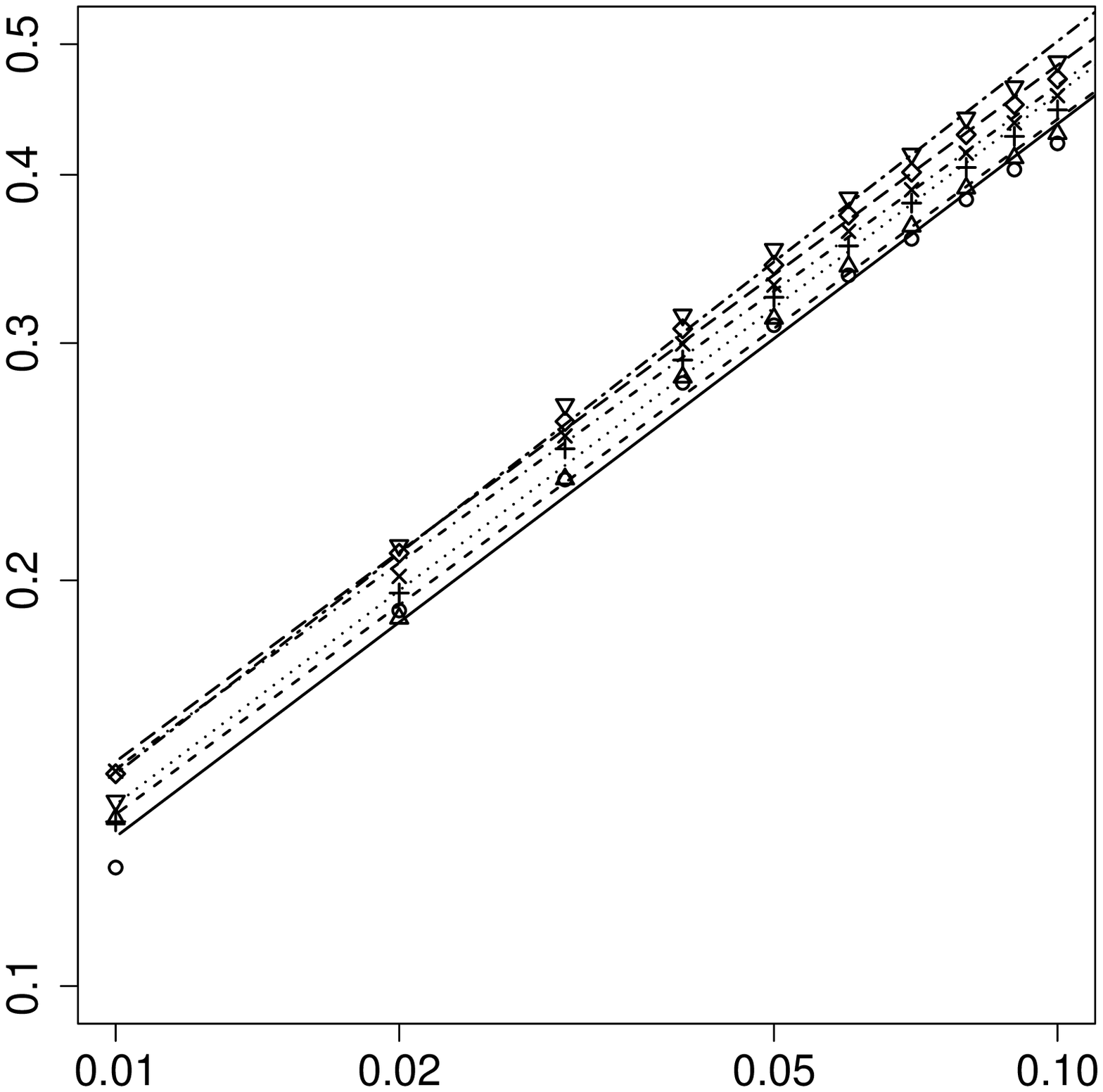}
\includegraphics[scale=0.21]{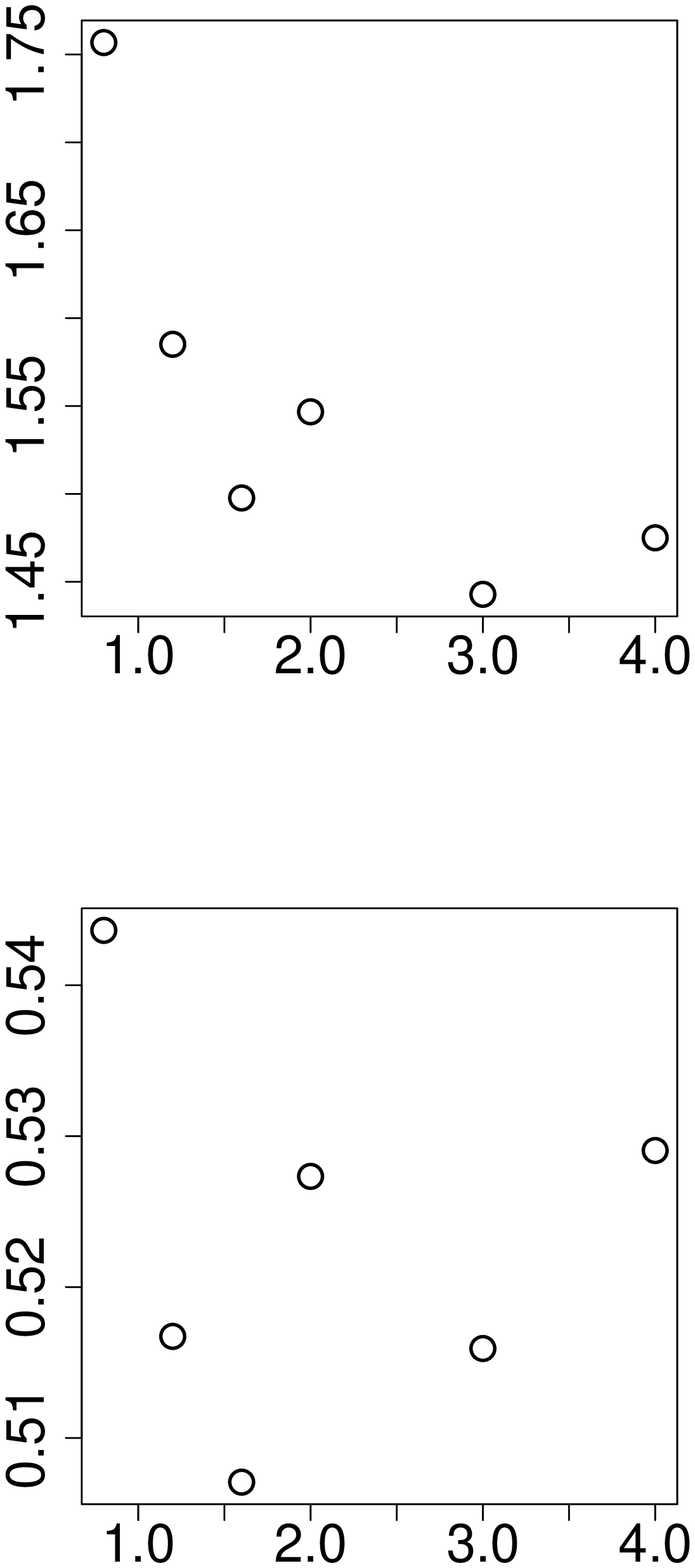}
\put(-83,29){$R_\infty$}
\put(-61,-3){$(K-K_c)/K_c$}
\put(-26,40){$\kappa$}
\put(-11,28){$\gamma$}
\put(-26,11){$\beta$}
\put(-11,-2){$\gamma$}
\caption{Critical behavior above the bifurcation using step $dK_r=0.01$.
Here, $\gamma$ is $4(\SA)$, $3(\SB)$, $2(\SC)$,
$1.6(\SD)$, $1.2(\SE)$ and $0.8(\SF)$.
Notice the log-log scale to emphasize the power law. Lines are fits to
$R_\infty=\kappa\,K_r^\beta$ Systems of sizes $N=10000$, $20000$ and
$50000$ were simulated with the same parameters and same scaling
procedure as in Fig.~\ref{kcfig}. Right panels show parameter
estimates from the left panel.}
\label{finer}
\end{figure}

Fig.~\ref{kcfig} fully supports the analytical solution and also indicates that
the behavior of $K_c$ above the bifurcation depends only weakly on $\ga$ over the simulated range.
To inspect more closely whether $R_\infty\sim\sqrt{K_r}$ as in the Kuramoto model,
we used refined spacing and larger sizes in the vicinity of $K_r=0^+$.
As shown in Fig.~\ref{finer}, the simulations are compatible with an
exponent $\beta=0.5$, the slightly higher exponents probably reflect a
finite size effect.
On the other hand $\kappa$ correlates negatively with $\gamma$ which
is visible in both Figs \ref{kcfig} and \ref{finer}.
%%%%%%%%%%%%%%%%%%%%%%%%%%%%%%%%%%%%%%%%
%%%%%%%%%%%%%%%%%%%%%%%%%%%%%%%%%%%%%%%
%%%%%%%%%%%%%%%%%%%%%%%%%%%%%%%%%%%%%%%%

\section{Discussion}

We have extended the Kuramoto model to frequencies which can drift
in time following Ornstein-Uhlenbeck dynamics.
The net effect is that the white noise source in Eq.~\ref{kuramoto} is
replaced by colored noise
(with a Cauchy distributed power spectrum), hereby
adding a new time scale describing memory or frequency stiffness to
the problem.
Apart from mean field coupling among the phases which introduces a
non-linearity, the stochastic phase and frequency dynamics lead to a linear
Fokker-Planck operator which can be solved.
Consequently the linear stability of the
incoherent state can be addressed analytically using
spectral calculus. The expression for the critical coupling above which
macroscopic phase coherence emerges can be resummed and expressed in
terms of incomplete $\Gamma$-functions. Asymptotic expansion for small
and large $\gamma$ shows that the full model continuously interpolates
between two limits of the original Kuramoto model: one dominated by
noise (large $\gamma$) and the other by the frequency dispersion
(small $\gamma$).
Therefore, the coupling force must counteract different sources of
fluctuations to induce collective synchrony in drifting frequency
oscillators, depending on the regime set by $\gamma$.
Specifically, for slowly drifting (small $\gamma$) frequencies
the model approaches the noiseless model (Eq.~\ref{kuramoto} with
$D=0$ and $g(f)=\NN_{\mu_f,\sigf}(f)$)
where the coupling splits the population into distinct locked and
incoherent sub-populations, depending
on the proximity of individual frequencies to the population mode.
As $\gamma$ increases, (while $\sigma_f$ is held fixed) the frequencies
lose their stiffness which results in a reduction in the critical
coupling $K_c$ needed for synchrony. Finally for very rapidly drifting
oscillators (large $\gamma$) cancel out the frequency distribution and
generate an effective white noise source acting on the phases of
otherwise equal frequency oscillators.
At the same time the locked and incoherent subgroups become indistinguishable.
For intermediate $\gamma$, our numerical simulations indicate
that the model belongs to the same $\beta=0.5$ universality class as the
Kuramoto model.

Because of analytical tractability and few parameters we expect this solution
to be relevant for oscillatory systems in the presence of complex noise sources.
Such cases include populations of neural oscillators or biochemical oscillators
where bioluminescence recordings have shown how intracellular noise sources
generate frequency dispersion through drifts.

\subsection*{Acknowledgments}
We thank Benoit Kornmann and Ueli Schibler for initiating our interest in
the drifting frequency model and Olivier Hernandez for pointing out useful
references. The simulations were performed on an Itanium2
cluster from HP/Intel at the Vital-IT facilities.
FN acknowledges funding from the Swiss National Science Foundation
NCCR Molecular Oncology program and NIH administrative supplement to parent
grant GM54339.

%\bibliography{rougemont}

\end{document}